\documentclass[twocolumn,prd,superscriptaddress,nofootinbib]{revtex4-2}

\usepackage{amsmath,amsfonts,amssymb}
\usepackage{bm}
\usepackage{graphicx}
\usepackage{xcolor}
\usepackage[utf8]{inputenc}
\usepackage{orcidlink}
\usepackage{hyperref}
\hypersetup{colorlinks=true,linkcolor=blue,citecolor=cyan,urlcolor=cyan}
\newcommand{\dd}{\mathrm{d}}
\newcommand{\ext}{\mathrm{ext}}

\begin{document}

\title{Schwinger stability of T-duality-inspired extremal black holes}

\author{Chiang-Mei Chen\,\orcidlink{0000-0003-2246-364X}}
\email{cmchen@phy.ncu.edu.tw}
\affiliation{Department of Physics, National Central University, Zhongli, Taoyuan 320317, Taiwan}
\affiliation{Center for High Energy and High Field Physics (CHiP), National Central University, Zhongli, Taoyuan 320317, Taiwan}

\author{Kimet Jusufi\,\orcidlink{0000-0003-0527-4177}}
\email{kimet.jusufi@unite.edu.mk}
\affiliation{Physics Department, State University of Tetovo, Ilinden Street nn, 1200 Tetovo, North Macedonia}

\author{Douglas Singleton\,\orcidlink{0000-0001-9155-7282}}
\email{dougs@mail.fresnostate.edu}
\affiliation{Physics Department, California State University, Fresno, California 93740, USA}

\begin{abstract}
We study the near-horizon charged-scalar instability associated with Schwinger pair production in the charged regular black-hole geometry of the zero-point-length T-duality prescription. Using the same effective geometry and regularized gauge potential, we construct the extremal branch, its ${\rm AdS}_2 \times S^2$ near-horizon throat, and the charged-scalar instability threshold for a general form factor. For the explicit T-duality form factor the extremal branch
terminates at $r_\mathrm{ext} = \sqrt2 \, l_0$, where the extremal charge and the near-horizon electric field vanish while the ${\rm AdS}_2$ radius remains finite at $\sqrt3 \, l_0$. At this endpoint the exact near-horizon instability parameter is negative. In the semiclassical regime the corresponding charge-to-mass threshold diverges as the endpoint is approached. Thus, for each fixed massive species with finite $q/m$, a sufficiently near-endpoint portion of the extremal branch is free of the local charged-scalar instability. For comparison, the Ay\'on–Beato–Garc\'ia (ABG) Einstein-nonlinear-electrodynamics solution has a nonvanishing extremal near-horizon electric field. For an additional minimally coupled charged-scalar probe, the corresponding local Schwinger threshold remains finite. Thus the divergent near-endpoint Schwinger barrier found in the T-duality branch is not a generic consequence of regularity.
\end{abstract}
\maketitle

\section{Introduction}

Near-extremal charged black holes develop an ${\rm AdS}_2 \times S^2$ throat, making them a useful laboratory for quantum fields in strong electric backgrounds. In the Reissner--Nordstr\"om (RN) geometry, charged particles can be produced by the Schwinger mechanism even as the Hawking temperature tends to zero at extremality~\cite{Chen:2012zn, Schwinger:1951nm, Gibbons:1975kk}. The near-horizon charged-scalar equation can be phrased as a violation of an effective
Breitenlohner--Freedman bound in ${\rm AdS}_2$~\cite{Chen:2014yfa, Cai:2020trh, Chen:2020mqs}.

A separate line of work motivated by string T-duality and zero-point length replaces pointlike short-distance behavior by an effective minimal scale $l_0$~\cite{Padmanabhan:1996ap, Fontanini:2005ik, Nicolini:2019irw}. For static gravitational and electric sources the prescription gives regularized potentials
\begin{equation}
V_G(r) = - \frac{m_0}{\sqrt{r^2 + l_0^2}},
\qquad
A_t(r) = -\frac{Q}{\sqrt{r^2 + l_0^2}}.
\label{eq:regular-potentials}
\end{equation}
The charged black-hole construction of Ref.~\cite{Gaete:2022ukm}, obtained within the corresponding static regularized-source prescription, contains a nontrivial radial form factor $\mathcal F(r)$ multiplying the charge contribution to the metric.  At radii of order $l_0$ this function differs appreciably from unity and therefore affects extremality and the near-horizon geometry.

The central point of the present work is that the physical electric potential remains the one in Eq.~\eqref{eq:regular-potentials}. It is this $A_t$ that determines the electric field, appears in the matter covariant derivative, and is used in the Schwinger calculation. The form factor $\mathcal F(r)$ is instead a property of the gravitational response to the regularized charge distribution. As we show in Sec.~\ref{sec:geometry}, $\mathcal F(r)$ should more precisely be viewed as packaging the radial contribution of the regularized charge sector after the mass function has been rewritten in terms of the asymptotic ADM mass. It is therefore related to a combination of the total electrostatic self-energy and the energy enclosed inside radius $r$, rather than simply to the enclosed electrostatic energy itself. In particular, it should not be confused with a redefinition of the gauge field.

In the local probe approximation used below, the charged scalar couples to the original regularized $A_t$, while the full $\mathcal F(r)$ enters through the background extremal geometry. We derive the extremal mass and charge, the ${\rm AdS}_2$ radius, and the horizon electric field for a general form factor, and then obtain the corresponding charged-scalar instability criterion.

For the T-duality form factor, the extremal branch continuously connected to the RN limit extends down to $r_\ext = \sqrt2\,l_0$. Along this branch the extremal charge and the near-horizon electric field vanish at the lower endpoint while the ${\rm AdS}_2$ radius remains finite. At the endpoint the exact instability parameter is negative, and for each fixed charged-scalar species the local instability is absent sufficiently close to this endpoint.

The paper is organized as follows: Section~\ref{sec:geometry} reviews the T-duality static source prescription, the physical gauge potential, and the effective charged metric with the full form factor, and identifies the structure of the mass function. Section~\ref{sec:extremal} derives the general extremal branch and near-horizon geometry. Section~\ref{sec:schwinger} studies the local charged-scalar instability. Section~\ref{sec:abg} contrasts this result with the ABG Einstein--nonlinear-electrodynamics solution, using its own electromagnetic field rather than the T-duality gauge potential. We summarize in Sec.~\ref{sec:conclusions}. Throughout the paper we use units where $G = c = \hbar = 1$.

\section{T-duality regularized charged geometry}
\label{sec:geometry}

In the zero-point-length prescription motivated by T-duality, the static propagator is modified in such a way that the gravitational and electrostatic potentials remain finite at the origin~\cite{Padmanabhan:1996ap, Fontanini:2005ik, Gaete:2022une}. For a point bare mass $m_0$, the regularized gravitational potential is~\cite{Nicolini:2019irw}
\begin{eqnarray}
V_G &=& - m_0 \int \frac{\dd^3k}{(2\pi)^3} \, \frac{l_0 \sqrt{k^2}}{k^2} K_1 \! \left( l_0 \sqrt{k^2} \right) \mathrm{e}^{i \vec{k} \cdot \vec{r}}
\nonumber\\
&=& - \frac{m_0}{\sqrt{r^2 + l_0^2}},
\label{eq:VG}
\end{eqnarray}
where $K_1$ is a modified Bessel function of the second kind. Applying the flat spatial Laplacian used in this static smearing prescription gives
\begin{equation}
\rho^{\rm bare}(r) = \frac{1}{4\pi} \bar{\nabla}^2 V_G(r) = \frac{3 l_0^2 m_0}{4 \pi (r^2 + l_0^2)^{5/2}}.
\label{density}
\end{equation}
Here $\bar{\nabla}^2$ belongs to the effective static source construction; Eq.~\eqref{density} should not be interpreted as having been obtained by varying a fully covariant nonlocal gravitational action. The corresponding regularized electrostatic potential is~\cite{Gaete:2022ukm, Gaete:2022une}
\begin{equation}
A_t(r) = - \frac{Q}{\sqrt{r^2 + l_0^2}}.
\label{eq:At}
\end{equation}
Inserting the resulting electric field into the standard Maxwell stress tensor gives~\cite{Gaete:2022ukm}
\begin{equation}
\rho^{\rm EM}(r) = \frac{Q^2 r^2}{8 \pi (r^2 + l_0^2)^3}.
\label{eq:rhoEMtree}
\end{equation}
The standard Maxwell tensor constructed from this field alone is not separately conserved because the regularized electric field is sourced by the nonzero smeared current:
\begin{equation}
\nabla_\mu \tau^{\mu}{}_{\nu}{}^{\rm Max} = - F_{\nu\lambda} \mathfrak{J}^{\lambda}.
\label{eq:EM-nonconservation}
\end{equation}
Here $\tau^\mu{}_\nu{}^{\rm Max}$ is the standard Maxwell stress tensor and $\mathfrak J^\mu$ denotes the nonzero smeared current defined by $\nabla_\mu F^{\mu\nu} = 4 \pi \mathfrak J^\nu$, where the Maxwell equation yields the smeared charge density
\begin{equation}
\mathfrak{J}^t(r) = \frac{3 l_0^2 Q}{4 \pi (r^2 + l_0^2)^{5/2}}.
\label{eq:Jt_smeared}
\end{equation}

Consequently, the Maxwell tensor alone does not constitute a complete conserved self-gravitating source for the metric below. In what follows, we use the metric obtained from the static source prescription of Ref.~\cite{Gaete:2022ukm} as an effective black-hole geometry; we do not claim here to derive its complete covariant stress tensor from an underlying T-duality action.

We take the effective static, spherically symmetric geometry in Schwarzschild gauge to be
\begin{equation}
\dd s^2 = -f(r) \dd t^2 + \frac{\dd r^2}{f(r)} + r^2 \dd\Omega^2,
\quad
f(r) = 1 - \frac{2 m(r)}{r}.
\label{eq:metric-ansatz}
\end{equation}
Following the static source prescription of~\cite{Gaete:2022ukm}, we use the regularized Maxwell density~\eqref{eq:rhoEMtree} as the charge-sector contribution to the static mass function of the form
\begin{equation}
m(r) = \frac{m_0 r^3}{(r^2 + l_0^2)^{3/2}} + m_Q(r),
\label{eq:massfunction}
\end{equation}
where
\begin{equation}
m_Q(r) = \frac{Q^2}{2}  \int_0^r  \frac{s^4\,\dd s}{(s^2 + l_0^2)^3}.
\label{eq:mQ}
\end{equation}
Carrying out the integral and using Eq.~\eqref{eq:Mshift} reproduces the effective metric employed in Ref.~\cite{Gaete:2022ukm}.
\begin{equation}
f(r) = 1 - \frac{2 M r^2}{(r^2 + l_0^2)^{3/2}} + \frac{Q^2 r^2}{(r^2 + l_0^2)^2} \, \mathcal{F}(r),
\label{eq:metric-generalF}
\end{equation}
where
\begin{equation}
\mathcal{F} \!=\! \frac{5}{8} \!+\! \frac{3 l_0^2}{8 r^2} \!+\! \frac{3 \pi \sqrt{r^2 \!+\! l_0^2}}{16 l_0} \left[ 1 \!-\! \frac{2 (r^2 \!+\! l_0^2)^{3/2}}{\pi r^3} \arctan\!\left( \frac{r}{l_0} \right) \right].
\label{eq:F_r}
\end{equation}
The Maxwell density contributes a finite total energy,
\begin{equation}
E_Q^\infty \equiv 4 \pi \int_0^\infty \rho^{\rm EM}(r) \, r^2 \dd r = \frac{3 \pi Q^2}{32 l_0},
\label{eq:EQinf}
\end{equation}
so within this source prescription the asymptotic mass parameter is
\begin{equation}
M = m_0 + E_Q^\infty.
\label{eq:Mshift}
\end{equation}
It is useful to make explicit what the metric form factor represents. Defining
\begin{equation}
h(r) = \frac{r^3}{(r^2 + l_0^2)^{3/2}},
\end{equation}
the mass function can be rewritten as
\begin{equation}
m(r) = M h(r) - \left[ E_Q^\infty h(r) - m_Q(r) \right].
\label{eq:massfunction-ADM}
\end{equation}
Therefore the positive charge term displayed in Eq.~\eqref{eq:metric-generalF} corresponds to
\begin{equation}
\frac{Q^2 r^2}{(r^2 + l_0^2)^2} \mathcal{F}(r) = \frac{2}{r} \left[ E_Q^\infty h(r) - m_Q(r) \right].
\label{eq:F-energy-combination}
\end{equation}
Thus $\mathcal{F}(r)$ is not simply the electrostatic energy enclosed inside $r$; it packages the radial contribution of the regularized charge sector after the mass function is expressed in terms of the ADM mass $M$. Its limiting values are $\mathcal{F}(0) = 3 \pi/16$ and $\mathcal{F}(r) \to 1$ as $r \to \infty$, so the metric approaches RN at large radius.

\section{Extremal branch with the full form factor}
\label{sec:extremal}

\subsection{Extremal mass and charge}

We now retain the full form factor ${\mathcal F}$ of the effective geometry and solve the extremality conditions directly. Introduce dimensionless variables
\begin{equation}
x = \frac{r}{l_0}, \qquad \alpha=\frac{r_\ext}{l_0}, \qquad \bar{M} = \frac{M}{l_0}, \qquad \bar{Q} = \frac{Q}{l_0},
\end{equation}
and write $F(x) \equiv \mathcal{F}(l_0 x)$, that is
\begin{equation}
F(x) = \frac58 + \frac{3}{8 x^2} + \frac{3 \pi \sqrt{1 \!+\! x^2}}{16} \left[ 1 \!-\! \frac{(1 \!+\! x^2)^{3/2}}{x^3} \frac{2}{\pi} \arctan x \right].
\label{eq:F-x}
\end{equation}
Then
\begin{equation}
f(x) = 1 - 2 \bar{M} \, a(x) + \bar{Q}^2 b(x),
\end{equation}
where
\begin{equation}
a(x) = \frac{x^2}{(1 + x^2)^{3/2}}, \qquad b(x) = \frac{x^2 F(x)}{(1 + x^2)^2}.
\end{equation}
An extremal horizon satisfies
\begin{equation}
f(\alpha) = 0, \qquad \frac{\dd f}{\dd x}(\alpha) = 0.
\label{eq:extremal-conditions}
\end{equation}
Defining
\begin{equation}
D(\alpha) = \alpha F(\alpha) - (1 + \alpha^2) F'(\alpha),
\label{eq:Dalpha}
\end{equation}
where the prime denotes differentiation with respect to the dimensionless argument, the solution of Eq.~\eqref{eq:extremal-conditions} is
\begin{equation}
\bar{Q}_{\ext}^{\,2} = \frac{(\alpha^2 - 2) (1 + \alpha^2)^2}{\alpha^3 D(\alpha)}
\label{eq:Qext-general}
\end{equation}
and
\begin{equation}
\bar M_{\ext} = \frac{(1 + \alpha^2)^{3/2} \left[ 2 (\alpha^2 - 1) F - \alpha (1 + \alpha^2) F' \right]}{2 \alpha^3 D(\alpha)}.
\label{eq:Mext-general}
\end{equation}

For the explicit function~\eqref{eq:F-x}, $D(\sqrt2) \simeq 0.7196412 > 0$. Numerical evaluation shows that $D(\alpha) > 0$ along the extremal branch continuously connected to the RN limit, including the interval $\alpha \geq \sqrt2$. On this branch Eq.~\eqref{eq:Qext-general} gives a real nonzero charge for $\alpha > \sqrt2$, while
\begin{equation}
\bar Q_{\ext} \to 0 \qquad \left( \alpha \to \sqrt2^+ \right).
\label{eq:Qzero}
\end{equation}
We therefore identify $\alpha = \sqrt2$ as the lower endpoint of this RN-connected extremal branch. At the endpoint the charge term disappears and the neutral extremal
mass is
\begin{equation}
M_{\ext}(\sqrt2) = \frac{3 \sqrt3}{4} l_0 \simeq 1.29904 \, l_0,
\label{eq:Mneutral}
\end{equation}
the value already fixed by the smeared mass profile alone. The lower end of the extremal branch therefore survives the full $\mathcal{F}$ deformation.

\subsection{Near-horizon geometry}

Near a degenerate horizon,
\begin{equation}
f(r) = \tfrac12 f''(r_\ext) (r - r_\ext)^2 + \mathcal{O}\left[ (r - r_\ext)^3 \right],
\end{equation}
and we define
\begin{equation}
L_2^2 = \frac{2}{f''(r_\ext)}.
\label{eq:L2def}
\end{equation}
With $r = r_\ext + \epsilon \rho$ and $t = \tau/\epsilon$, followed by $\epsilon \to 0$, the metric becomes
\begin{equation}
\dd s^2 = - \frac{\rho^2}{L_2^2} \dd\tau^2 + \frac{L_2^2}{\rho^2} \dd\rho^2 + r_\ext^2 \dd\Omega^2,
\label{eq:AdS2S2}
\end{equation}
that is, ${\rm AdS}_2(L_2) \times S^2(r_\ext)$.

For compactness define
\begin{eqnarray}
N(\alpha) &=& 2 \alpha (\alpha^4 - \alpha^2 + 4) F
- (2 \alpha^6 + \alpha^4 + 5 \alpha^2 + 6) F'
\nonumber\\
&+& \alpha (\alpha^6 - 3 \alpha^2 - 2) F''.
\label{eq:Nalpha}
\end{eqnarray}
Direct evaluation of $f''$ on the extremal branch gives
\begin{equation}
\frac{L_2^2}{l_0^2} = \frac{2 \alpha^2 (1 + \alpha^2)^2 D(\alpha)}{N(\alpha)}.
\label{eq:L2-general}
\end{equation}
For the explicit T-duality form factor, $N\left( \sqrt2 \right) \simeq 8.6356938$, and Eq.~\eqref{eq:L2-general} gives the limiting value
\begin{equation}
L_2^2 \to 3 l_0^2, \qquad L_2 \to \sqrt3\,l_0.
\label{eq:L2-limit}
\end{equation}
The limiting throat is therefore ${\rm AdS}_2(\sqrt3\,l_0) \times S^2(\sqrt2\,l_0)$, even though $F(\sqrt2) \neq 1$: the deformation cancels between $D$ and $N$ at the endpoint.

\subsection{Near-horizon electric field}

The gauge potential remains Eq.~\eqref{eq:At}.  Subtracting its constant horizon value and taking the near-horizon scaling gives
\begin{equation}
A_\tau = E_H \rho, \qquad E_H = \frac{Q_{\ext} r_{\ext}}{(r_{\ext}^2 + l_0^2)^{3/2}},
\label{eq:EH}
\end{equation}
and combining with Eq.~\eqref{eq:Qext-general},
\begin{equation}
(E_H l_0)^2 = \frac{\alpha^2 - 2}{\alpha (1 + \alpha^2) D(\alpha)}.
\label{eq:EH-general}
\end{equation}
Since $D(\sqrt2)$ is finite and positive,
\begin{equation}
E_H \to 0 \qquad \left( \alpha \to \sqrt2^+ \right),
\label{eq:EH-zero}
\end{equation}
with the near-endpoint behavior
\begin{equation}
E_H l_0 \simeq 0.57230 \sqrt{\alpha^2 - 2}.
\label{eq:EH-near}
\end{equation}
The vanishing of the electric driving field, while $L_2$ stays finite, is the central geometric input to the Schwinger analysis.

\section{Near-Horizon Schwinger Instability}
\label{sec:schwinger}

We now use the same regularized gauge field used in the effective black-hole construction, Eq.~\eqref{eq:At}, and couple it to a scalar field, $\Psi$, giving
\begin{equation}
(D_\mu D^\mu - m^2) \Psi = 0, \qquad D_\mu = \nabla_\mu - i q A_\mu.
\label{eq:KG}
\end{equation}
In this probe approximation, the zero-point length is included only through the background geometry and electric field; no minimum-length modification of the scalar propagator is assumed. The form factor modifies the extremal geometry and hence the near-horizon values of $Q_\ext$ and $L_2$, whereas the electric driving force is still obtained by differentiating the physical $A_t$.  This is a semiclassical probe calculation, not a first-principles T-duality pair-production rate.

Using
\begin{equation}
\Psi = \mathrm{e}^{-i \omega \tau} R(\rho) Y_{\ell m}(\theta, \varphi),
\end{equation}
the radial equation in the throat~\eqref{eq:AdS2S2} takes the form
\begin{eqnarray}
\frac{\dd}{\dd\rho} \left( \rho^2 \frac{\dd R}{\dd\rho} \right) &+& \left[ \frac{L_2^4 (\omega + q E_H \rho)^2}{\rho^2} - m^2 L_2^2 \right.
\nonumber\\
&& \left. - \ell (\ell + 1) \frac{L_2^2}{r_{\ext}^2} \right] R = 0~.
\label{eq:radial-KG}
\end{eqnarray}
From~\eqref{eq:radial-KG} one can define an effective instability parameter
\begin{equation}
\mu_*^2 = (q E_H L_2^2)^2 - m^2 L_2^2 - \ell (\ell + 1) \frac{L_2^2}{r_{\ext}^2} - \frac14.
\label{eq:mustar}
\end{equation}
The near-horizon charged-scalar instability occurs when
\begin{equation}
\mu_*^2 > 0,
\end{equation}
corresponding to violation of the effective ${\rm AdS}_2$ Breitenlohner--Freedman bound. At the lower endpoint, the exact instability parameter is manifestly negative.  Using $E_H \to 0$, $L_2^2 \to 3 l_0^2$, and $r_{\ext}^2 \to 2 l_0^2$, Eq.~\eqref{eq:mustar} gives
\begin{equation}
\mu_*^2 \longrightarrow - 3 m^2 l_0^2 - \frac{3}{2} \ell (\ell + 1) - \frac14 < 0.
\end{equation}
Thus the local charged-scalar instability is absent at the endpoint. Since $\mu_*^2(\alpha)$ is continuous along the extremal branch near $\alpha = \sqrt2$, for each fixed $q$, $m$, and $\ell$ there is a sufficiently near-endpoint neighborhood in which $\mu_*^2 < 0$. This conclusion does not rely on the semiclassical approximation used below.

For the most favorable $\ell = 0$ mode and in the semiclassical regime $m L_2 \gg 1$, the $1/4$ term can be neglected and the condition reduces to $|q E_H| L_2 > m$.  It is convenient to define
\begin{equation}
\frac{q^2}{m^2} > \Xi_{\mathcal F}(\alpha), \qquad \Xi_{\mathcal F}(\alpha) = \frac{1}{E_H^2 L_2^2}.
\label{eq:Xi-def}
\end{equation}
Using Eqs.~\eqref{eq:L2-general} and~\eqref{eq:EH-general}, all factors of $D$ cancel and
\begin{equation}
\Xi_{\mathcal F}(\alpha) = \frac{N(\alpha)}{2 \alpha (\alpha^2 - 2) (1 + \alpha^2)},
\label{eq:Xi-general}
\end{equation}
or explicitly
\begin{eqnarray}
\Xi_{\mathcal F} &=& \frac{1}{2 \alpha (\alpha^2 - 2) (1 + \alpha^2)} \Big\{ 2 \alpha (\alpha^4 - \alpha^2 + 4) F
\nonumber\\
&-& (2 \alpha^6 + \alpha^4 + 5 \alpha^2 + 6) F'
\nonumber\\
&+& \alpha (\alpha^6 - 3 \alpha^2 - 2) F'' \Big\}.
\label{eq:Xi-expanded}
\end{eqnarray}
As $\alpha \to \infty$, $F \to 1$ and its derivatives vanish, so $\Xi_{\mathcal F} \to 1$, recovering the extremal RN threshold. Most importantly, close to the lower end of the extremal branch,
\begin{equation}
\Xi_{\mathcal F}(\alpha) \simeq \frac{1.017726}{\alpha^2 - 2} \longrightarrow \infty.
\label{eq:Xi-diverge}
\end{equation}
In the semiclassical $\ell = 0$ regime, Eq.~\eqref{eq:Xi-diverge} quantifies this approach to the endpoint: for each fixed finite $q/m$, the instability condition~\eqref{eq:Xi-def} eventually fails as $\alpha \to \sqrt2^+$. The divergence is the direct counterpart of the vanishing $E_H$ in Eq.~\eqref{eq:EH-zero}: the ${\rm AdS}_2$ radius remains finite and therefore cannot compensate for the disappearing electric field.

\section{Comparison with the Ay\'on--Beato--Garc\'ia black hole}
\label{sec:abg}

The ABG regular black hole provides a useful benchmark for the near-horizon mechanism found above, but it should not be regarded as a parameter limit of the T-duality solution. At the level of the metric, Eq.~\eqref{eq:metric-generalF} can be made algebraically identical to the ABG metric by the two independent replacements
\begin{equation}
l_0 \longrightarrow |Q|, \qquad \mathcal{F}(r) \longrightarrow 1.
\label{eq:ABG-metric-replacements}
\end{equation}
Setting $l_0 = |Q|$ alone merely changes the argument of the nontrivial T-duality form factor to $\mathcal{F}(r/|Q|)$ and does not set it to unity. Hence the simultaneous replacements in Eq.~\eqref{eq:ABG-metric-replacements} change the source model rather than define a finite-$Q$ limit of the T-duality family. The resulting metric is
\begin{equation}
f_{\rm ABG}(r) = 1 - \frac{2 M r^2}{(r^2 + Q^2)^{3/2}} + \frac{Q^2 r^2}{(r^2 + Q^2)^2},
\label{eq:ABG-metric}
\end{equation}
which is the Ay\'on--Beato--Garc\'ia geometry~\cite{Ayon-Beato:1998hmi}. The complete ABG solution, however, is an Einstein--nonlinear-electrodynamics (NLED) system and therefore carries its own electromagnetic field. It is not consistent to combine Eq.~\eqref{eq:ABG-metric} with the T-duality potential Eq.~\eqref{eq:At}; doing so would define a hybrid model rather than the ABG black hole.

For the ABG solution the electric field supporting the geometry is
\begin{equation}
E_{\rm ABG}(r) = Q r^4 \left[ \frac{r^2 - 5 Q^2}{(r^2 + Q^2)^4} + \frac{15 M}{2 (r^2 + Q^2)^{7/2}} \right].
\label{eq:ABG-electric-field}
\end{equation}
We therefore redo the local near-horizon charged-scalar analysis using Eq.~\eqref{eq:ABG-electric-field}.  Introducing $\alpha = r_{\rm ext}/|Q|$, the extremality conditions $f_{\rm ABG}(r_{\rm ext}) = f'_{\rm ABG}(r_{\rm ext}) = 0$ give
\begin{equation}
(\alpha^2 - 2) (1 + \alpha^2)^2 = \alpha^4,
\label{eq:ABG-cubic-alpha}
\end{equation}
or, with $y = \alpha^2$,
\begin{equation}
y^3 - y^2 - 3 y - 2 = 0.
\label{eq:ABG-cubic}
\end{equation}
The physical root and the associated extremal mass are
\begin{equation}
\alpha_{\rm ABG} \simeq 1.584786, \qquad \frac{M_{\rm ext}}{|Q|} \simeq 1.57684.
\label{eq:ABG-extremal-values}
\end{equation}
Thus the extremal ABG solution occurs at a fixed dimensionless radius; it is not a trajectory obtained by driving an independent T-duality scale toward the endpoint $r_{\rm ext} = \sqrt2 \, l_0$.

Expanding the ABG metric around its degenerate horizon gives the standard ${\rm AdS}_2 \times S^2$ throat with
\begin{equation}
L_{2,{\rm ABG}}^2 = \frac{2}{f''_{\rm ABG}(r_{\rm ext})}, \qquad \frac{L_{2,{\rm ABG}}}{|Q|} \simeq 1.9931.
\label{eq:ABG-L2}
\end{equation}
Evaluating the NLED field in Eq.~\eqref{eq:ABG-electric-field} at the same extremal horizon gives
\begin{equation}
|E_H^{\rm ABG}| \, |Q| \simeq 0.8161.
\label{eq:ABG-EH}
\end{equation}
The absolute value is displayed because the sign of the electric field follows the sign of $Q$, whereas the local Schwinger criterion depends on $E_H^2$. In particular, the ABG extremal electric field remains finite and nonzero, in sharp contrast with the T-duality branch where $E_H \rightarrow 0$ as $\alpha \rightarrow \sqrt2^+$.

We assume that an additional scalar, charged under the same $U(1)$, couples minimally to the ABG gauge potential; this is a test-field assumption rather than part of the original ABG matter model. Its exact near-horizon instability parameter then has the same kinematical form as Eq.~\eqref{eq:mustar}, but with the ABG throat and ABG NLED field,
\begin{eqnarray}
(\mu_*^{\rm ABG})^2 &=& (q E_H^{\rm ABG} L_{2,{\rm ABG}}^2)^2 - m^2 L_{2,{\rm ABG}}^2
\nonumber\\
&-& \ell (\ell + 1) \frac{L_{2,{\rm ABG}}^2}{r_{\rm ext}^2} - \frac14.
\label{eq:ABG-mustar}
\end{eqnarray}
For the most favorable $\ell = 0$ mode this can be written exactly as
\begin{equation}
\frac{q^2}{m^2} > \frac{1}{(E_H^{\rm ABG})^2 L_{2,{\rm ABG}}^2} \left( 1 + \frac{1}{4 m^2 L_{2,{\rm ABG}}^2} \right).
\label{eq:ABG-exact-threshold}
\end{equation}
In the same semiclassical regime $m L_{2,{\rm ABG}} \gg 1$ used in Sec.~\ref{sec:schwinger}, the $1/4$ term is negligible and the local criterion reduces to
\begin{equation}
\frac{q^2}{m^2} > \Xi_{\rm ABG}, \quad \Xi_{\rm ABG} = \frac{1}{(E_H^{\rm ABG})^2 L_{2,{\rm ABG}}^2} \simeq 0.378,
\label{eq:ABG-Xi}
\end{equation}
so that
\begin{equation}
\left( \frac{|q|}{m} \right)_{\rm crit} \simeq 0.615.
\label{eq:ABG-threshold}
\end{equation}
The contrast with the T-duality branch can therefore be summarized as
\[
\begin{aligned}
\text{T-duality branch:} \quad & E_H l_0 \longrightarrow 0, & \Xi_{\mathcal F} \longrightarrow \infty,
\\
\text{extremal ABG:} \quad & |E_H^{\rm ABG}| \, |Q| \simeq 0.816, & \Xi_{\rm ABG} \simeq 0.378.
\end{aligned}
\]
Thus regularity alone does not imply a divergent Schwinger barrier; the suppression arises from the specific T-duality extremal branch on which the physical horizon electric field vanishes. Probe species with sufficiently large $|q|/m$ can therefore satisfy the local near-horizon Schwinger-instability criterion.

This local Schwinger instability should not be confused with a proof of complete black-hole evaporation.  An exactly extremal ABG black hole has $T_H = 0$, and once charged emission occurs the parameters $M$ and $Q$ evolve away from the original extremal point.  Establishing the ultimate endpoint requires the coupled Hawking and Schwinger fluxes, greybody factors, charge quantization, and backreaction.

Independently, the NLED sector has a classical stability issue.  De Felice and Tsujikawa showed that nonsingular black holes realized within nonlinear electrodynamics are subject to an angular Laplacian instability around the regular center, associated with a negative squared angular propagation speed of a vector perturbation and its coupling to the gravitational even-parity sector~\cite{DeFelice:2025nled}. Since ABG is a nonsingular Einstein–NLED black hole, it belongs to the class of solutions addressed by Ref.~\cite{DeFelice:2025nled}.  It is conceptually distinct from the Schwinger effect, however: the former concerns perturbations of the coupled background itself, whereas the latter concerns an additional charged quantum field propagating on that background.  A dedicated mode analysis of the specific ABG NLED model lies beyond the present comparison.

\section{Conclusions}
\label{sec:conclusions}

We have analyzed the charged zero-point-length black hole while retaining the full T-duality form factor in the effective metric. The same regularized electrostatic potential is used consistently at all stages: it generates the electric field entering the original source construction, and couples to the charged scalar in the Schwinger calculation.  The form factor is not a modification of this gauge potential.  As Eqs.~\eqref{eq:massfunction-ADM} and~\eqref{eq:F-energy-combination} show, it packages the radial contribution of the regularized charge sector after the mass function is rewritten in terms of the ADM mass.  It is therefore not simply the electrostatic energy enclosed within radius $r$, and it belongs to the mass-function sector rather than to $A_\mu$.

The extremal and near-horizon analysis was then carried out on this same background.  The extremal mass, charge, and ${\rm AdS}_2$ radius depend on the form factor and its derivatives.  For the explicit T-duality function the branch nevertheless reaches the limiting radius $r_{\ext} = \sqrt2 \, l_0$, with the extremal charge and the physical near-horizon electric field tending to zero while the ${\rm AdS}_2$ radius approaches the finite value $\sqrt3 \, l_0$.  This limiting behavior is therefore a property of the full form factor of the effective geometry rather than of a simplified charge sector.

Using the original gauge potential in the charged-scalar covariant derivative, we derived the generalized Schwinger/Breitenlohner--Freedman instability criterion.  At the endpoint itself the exact instability parameter is negative, so the absence of the local charged-scalar instability there does not rely on the semiclassical $m L_2 \gg 1$ approximation.  By continuity, each fixed charged-scalar species has a sufficiently near-endpoint neighborhood in which the local instability remains absent. In the semiclassical $\ell = 0$ regime, the divergence $\Xi_{\mathcal F} \to \infty$ describes how this suppression is approached along the extremal branch.

It is also instructive to compare the present construction with the ABG Einstein--NLED black hole.  The ABG metric can be matched algebraically by the two independent replacements $l_0 \to |Q|$ and $\mathcal{F} \to1$, but this is not a parameter limit of the T-duality source model.  Using the actual ABG NLED electric field, the extremal solution has $r_{\rm ext}/|Q| \simeq 1.585$, $L_2/|Q| \simeq 1.993$, and $|E_H^{\rm ABG}|\,|Q| \simeq 0.816$, with $|Q_{\rm ext}|/M_{\rm ext} \simeq 0.6342$. The corresponding semiclassical $s$-wave threshold is finite, $\Xi_{\rm ABG} \simeq 0.378$, or $(|q|/m)_{\rm crit} \simeq 0.615$, rather than the value obtained from the T-duality gauge field.  Hence the divergent Schwinger barrier is not a generic consequence of regularity: it is tied to the specific T-duality extremal branch on which the physical horizon electric field vanishes.  This local comparison does not establish complete evaporation of ABG.  Moreover, the independent NLED perturbative instability identified by De Felice and Tsujikawa~\cite{DeFelice:2025nled} is a distinct classical issue and should not be conflated with charged-particle Schwinger production.

The T-duality Schwinger interpretation has two additional limitations: (i) The produced matter field has been treated with an ordinary local propagator; a microscopic minimum-length theory may modify the pair-production criterion. (ii) The existence of a Schwinger-suppressed extremal region does not prove that an evaporating black hole reaches it dynamically.  Establishing an actual endpoint requires the coupled evolution of $M(t)$ and $Q(t)$, including Hawking and Schwinger fluxes, greybody factors, and backreaction.

These qualifications do not affect the geometric endpoint result: the horizon electric field vanishes while the ${\rm AdS}_2$ radius remains finite. Within the local charged-scalar probe model adopted here, this makes the exact instability parameter negative and eliminates the near-horizon charged-scalar instability at the endpoint.

\bigskip
\begin{acknowledgments}
C.M.C. would like to thank the Department of Physics and Kobayashi-Maskawa Institute, Nagoya University, and the Department of Physics, Rikkyo University, for hospitality while part of this work was done. The work of C.M.C. was supported by the National Science and Technology Council of the R.O.C. (Taiwan) under grants NSTC 115-2112-M-008-002 and 115-2918-I-008-006.
D.S. acknowledges the support of the Frank Sutton Research Fund.
\end{acknowledgments}

\bibliography{references}

@PREAMBLE{
 "\providecommand{\noopsort}[1]{}" 
 # "\providecommand{\singleletter}[1]{#1}%" 
}

@article{Ayon-Beato:1998hmi,
    author = "Ayon-Beato, Eloy and Garcia, Alberto",
    title = "{Regular black hole in general relativity coupled to nonlinear electrodynamics}",
    eprint = "gr-qc/9911046",
    archivePrefix = "arXiv",
    doi = "10.1103/PhysRevLett.80.5056",
    journal = "Phys. Rev. Lett.",
    volume = "80",
    pages = "5056--5059",
    year = "1998"
}

@article{Gaete:2022ukm,
    author = "Gaete, Patricio and Jusufi, Kimet and Nicolini, Piero",
    title = "{Charged black holes from T-duality}",
    eprint = "2205.15441",
    archivePrefix = "arXiv",
    primaryClass = "hep-th",
    doi = "10.1016/j.physletb.2022.137546",
    journal = "Phys. Lett. B",
    volume = "835",
    pages = "137546",
    year = "2022"
}

@article{Cai:2020trh,
    author = "Cai, Rong-Gen and Chen, Chiang-Mei and Kim, Sang Pyo and Sun, Jia-Rui",
    title = "{Schwinger effect in near-extremal charged black holes in high dimensions}",
    eprint = "2004.00735",
    archivePrefix = "arXiv",
    primaryClass = "hep-th",
    doi = "10.1103/PhysRevD.101.105015",
    journal = "Phys. Rev. D",
    volume = "101",
    number = "10",
    pages = "105015",
    year = "2020"
}

@article{Chen:2020mqs,
    author = "Chen, Chiang-Mei and Kim, Sang Pyo",
    title = "{Schwinger Effect from Near-extremal Black Holes in (A)dS Space}",
    eprint = "2002.00394",
    archivePrefix = "arXiv",
    primaryClass = "hep-th",
    doi = "10.1103/PhysRevD.101.085014",
    journal = "Phys. Rev. D",
    volume = "101",
    number = "8",
    pages = "085014",
    year = "2020"
}

@article{Padmanabhan:1996ap,
    author = "Padmanabhan, T.",
    title = "{Duality and zero point length of space-time}",
    eprint = "hep-th/9608182",
    archivePrefix = "arXiv",
    reportNumber = "IUCAA-35-96",
    doi = "10.1103/PhysRevLett.78.1854",
    journal = "Phys. Rev. Lett.",
    volume = "78",
    pages = "1854--1857",
    year = "1997"
}

@article{Fontanini:2005ik,
    author = "Fontanini, Michele and Spallucci, Euro and Padmanabhan, T.",
    title = "{Zero-point length from string fluctuations}",
    eprint = "hep-th/0509090",
    archivePrefix = "arXiv",
    doi = "10.1016/j.physletb.2005.12.039",
    journal = "Phys. Lett. B",
    volume = "633",
    pages = "627--630",
    year = "2006"
}

@article{DeFelice:2025nled,
  author  = {De Felice, Antonio and Tsujikawa, Shinji},
  title   = {Instability of Nonsingular Black Holes in Nonlinear Electrodynamics},
  journal = {Phys. Rev. Lett.},
  volume  = {134},
  pages   = {081401},
  year    = {2025},
  doi     = {10.1103/PhysRevLett.134.081401}
}

@article{Nicolini:2019irw,
    author = "Nicolini, Piero and Spallucci, Euro and Wondrak, Michael F.",
    title = "{Quantum Corrected Black Holes from String T-Duality}",
    eprint = "1902.11242",
    archivePrefix = "arXiv",
    primaryClass = "gr-qc",
    doi = "10.1016/j.physletb.2019.134888",
    journal = "Phys. Lett. B",
    volume = "797",
    pages = "134888",
    year = "2019"
}

@article{Gaete:2022une,
    author = "Gaete, Patricio and Nicolini, Piero",
    title = "{Finite electrodynamics from T-duality}",
    eprint = "2202.09311",
    archivePrefix = "arXiv",
    primaryClass = "hep-th",
    doi = "10.1016/j.physletb.2022.137100",
    journal = "Phys. Lett. B",
    volume = "829",
    pages = "137100",
    year = "2022"
}

@article{Chen:2012zn,
    author = "Chen, Chiang-Mei and Kim, Sang Pyo and Lin, I-Chieh and Sun, Jia-Rui and Wu, Ming-Fan",
    title = "{Spontaneous Pair Production in Reissner-Nordstrom Black Holes}",
    eprint = "1202.3224",
    archivePrefix = "arXiv",
    primaryClass = "hep-th",
    doi = "10.1103/PhysRevD.85.124041",
    journal = "Phys. Rev. D",
    volume = "85",
    pages = "124041",
    year = "2012"
}

@article{Schwinger:1951nm,
    author = "Schwinger, Julian S.",
    editor = "Milton, K. A.",
    title = "{On gauge invariance and vacuum polarization}",
    doi = "10.1103/PhysRev.82.664",
    journal = "Phys. Rev.",
    volume = "82",
    pages = "664--679",
    year = "1951"
}

@article{Gibbons:1975kk,
    author = "Gibbons, G. W.",
    title = "{Vacuum Polarization and the Spontaneous Loss of Charge by Black Holes}",
    doi = "10.1007/BF01609829",
    journal = "Commun. Math. Phys.",
    volume = "44",
    pages = "245--264",
    year = "1975"
}

@article{Chen:2014yfa,
    author = "Chen, Chiang-Mei and Sun, Jia-Rui and Tang, Fu-Yi and Tsai, Ping-Yen",
    title = {{Spinor particle creation in near extremal Reissner\textendash{}Nordstr\"om black holes}},
    eprint = "1412.6876",
    archivePrefix = "arXiv",
    primaryClass = "hep-th",
    doi = "10.1088/0264-9381/32/19/195003",
    journal = "Class. Quant. Grav.",
    volume = "32",
    number = "19",
    pages = "195003",
    year = "2015"
}

\end{document}